# THE ORIGIN OF GRAVITY

C.P. Kouropoulos†

## *Abstract*

I consider a pair of harmonic oscillators coupled by the Coulomb and the electrokinetic potential, which decreases as *1/r*. Their dynamic coupling is allowed in the near field, that is, for frequencies roughly below ~ *c/r* for real or virtual exchanges within a larger coherence domain. Within a two-phase constraint, the correlated interacting modes have lower eigenfrequencies and positive zero-modes, their anticorrelating counterparts higher eigenfrequencies and negative zero-modes. While the former induce coherent states over large ensembles, the latter cannot. As a result, the cosmological constant depends only on the system's purely disordered phases. An attractive *–1/r* long-range potential is found owing to the coherent zero-modes of the near field in the far infrared. This can occur in Machian cosmologies such as in Yilmaz's, wherein the Zitterbewegung of very distant matter is felt in the far infrared by the Zitterbewegung of local matter, in proportion to the frequencies of both, in accordance with the principle of equivalence. The Machian inertia appears on cosmic scales from the coherent states through their tunneling radiation. In a closed elliptic Universe, the very slow modes of the coupled oscillators can be identified with those of their radiative antipodal image. Identifying the antipodes after a parity reversal, far infrared coherent modes arise in their local realm of the Universe, to which the local oscillators intrinsically partake, inducing gravity.

†e-mail : kouros@bluewin.ch

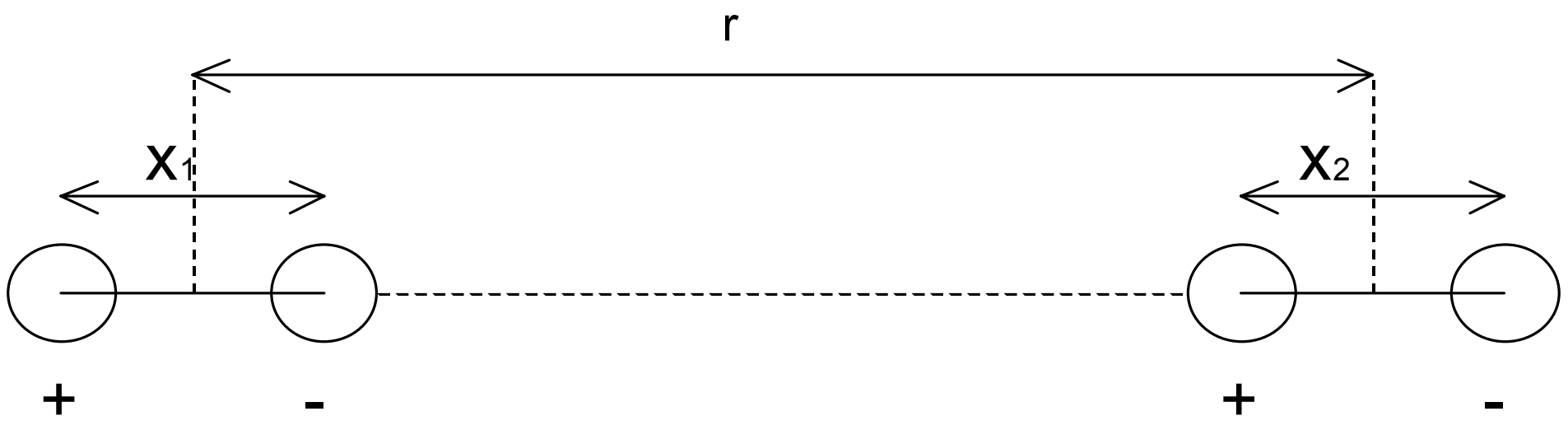

This represents two electric dipoles. Their mutual Coulomb coupling is

$$[1]\quad 4\pi\varepsilon_o U_{12} = -\frac{e^2}{r+\frac{x_1+x_2}{2}} - \frac{e^2}{r-\frac{x_1+x_2}{2}} + \frac{e^2}{r+\frac{x_1-x_2}{2}} + \frac{e^2}{r-\frac{x_1-x_2}{2}}$$

$$= \frac{e^2}{r}\left( -\frac{1}{1+\frac{x_1+x_2}{2r}} - \frac{1}{1-\frac{x_1+x_2}{2r}} + \frac{1}{1+\frac{x_2-x_1}{2r}} + \frac{1}{1-\frac{x_2-x_1}{2r}} \right) \quad ,$$

$$\frac{1}{1+\frac{a}{r}} = 1 - \frac{a}{r} + \frac{a^2}{r^2} - \ldots \quad ,$$

$$U_{12} = -\frac{1}{2\pi\varepsilon_o}\frac{e^2}{r^3}x_1 x_2 + \ldots$$

Besides the Coulomb coupling, there is the acceleration-dependent vector potential and the corresponding electrokinetic force

$$[2]\qquad \vec{A}_i = \frac{\mu_o}{2\pi}\frac{e^2 \dot{x}_j}{r}$$

$$F_{k1} = -\partial_t A_1 = -\frac{\mu_o}{2\pi}\frac{e^2 \ddot{x}_2}{r}$$

$$F_{k2} = -\partial_t A_2 = -\frac{\mu_o}{2\pi}\frac{e^2 \ddot{x}_1}{r} \quad ,$$

where we assumed that, in each oscillator, the acceleration of one charge opposes that of the other. The Lagrangian and the corresponding equations of motion are

$$[3] \quad L = \frac{1}{2} m\left(\dot{x}_1^2 + \dot{x}_2^2\right) - \frac{1}{2} m\omega_o^2\left(x_1^2 + x_2^2\right) + kx_1x_2 - k'\dot{x}_1\dot{x}_2$$

$$\ddot{x}_1 + \omega_o^2 x_1 - \frac{k}{m}x_2 + \frac{k'}{m}\ddot{x}_2 = 0$$

$$\ddot{x}_2 + \omega_o^2 x_2 - \frac{k}{m}x_1 + \frac{k'}{m}\ddot{x}_1 = 0 \quad .$$

with

$$[4] \quad \omega_o = \sqrt{\frac{f}{m}} \quad , \quad k = \frac{1}{2\pi\varepsilon_o}\frac{e^2}{r^3} \quad , \quad k' = \frac{\mu_o}{2\pi}\frac{e^2}{r} \quad .$$

The term $f$ is akin to a spring's tension and approximates the electrostatic and the electrokinetic attraction or repulsion between the neighbouring charges of one oscillator on a small displacement. We omitted the Lorentz force of the electrokinetic field from one dipole acting transversely to a speeding charge at $v/c=\beta$ on the other dipole, the magnitude of which is as $k'\beta /r$, as well as the purely magnetic Lorentz force $k'(\mathbf{v}_1 x\mathbf{r})x\mathbf{v}_2/r^3$. This manifestly requires our embedding the dipoles in to two or three dimensions. In a single dimension, taking

$$[5] \quad x_1 = Ae^{i\omega t} \quad , \quad x_2 = Be^{i\omega t} \quad ,$$

substituting and simplifying by $e^{i\omega t}$

$$[6] \quad A\left(\omega_o^2 - \omega^2\right) - \frac{B}{m}\left(k'\omega^2 + k\right) = 0$$

$$B\left(\omega_o^2 - \omega^2\right) - \frac{A}{m}\left(k'\omega^2 + k\right) = 0 \quad .$$

The canceling of the determinant of which gives the eigenfrequencies

[7] $$\begin{vmatrix} \omega_o^2-\omega^2 & \frac{1}{m}\left(k'\omega^2+k\right) \\ \frac{1}{m}\left(k'\omega^2+k\right) & \omega_o^2-\omega^2 \end{vmatrix} = 0$$

$\Rightarrow$

$$|\omega_\pm| = \omega_o\sqrt{\frac{1\pm\frac{k}{m\omega_o^2}}{1\mp\frac{k'}{m}}} \cong \omega_o\sqrt{1\pm\frac{1}{m}\left(k'+\frac{k}{\omega_o^2}\right)} \quad .$$

The solutions are $A = \pm B$. and correspond to correlated ($\omega_-$) and anticorrelated ($\omega_+$) motions. The Hamiltonian for the equations of motion [5] is

[8] $$T = \frac{1}{2m}\left(p_{x_1}{}^2 + p_{x_2}{}^2\right) \quad , \qquad U = \frac{m\omega_o^2}{2}\left(x_1^2+x_2^2\right) - kx_1x_2 \quad ,$$

$$p_i = m\dot{x}_i - k'\dot{x}_j \quad ,$$

$$H = T + U \quad ,$$

with

[9] $$\xi = \frac{1}{\sqrt{2}}(x_1+x_2) \quad , \quad \eta = \frac{1}{\sqrt{2}}(x_1-x_2) \quad ,$$

$$H = \frac{1}{2m}\left(p_\xi{}^2 + p_\eta{}^2\right) + \frac{m}{2}\left(\omega_-^2\xi^2 + \omega_+^2\eta^2\right) \quad ,$$

where $\omega_\pm$ are the frequencies [8]. The minimal energy is here related to the $\xi$ -states, which involve correlated motions. This is precisely the Hamiltonian for two distinct harmonic oscillators with the frequencies

$$[10] \quad \omega_{\pm} \cong \omega_o \left( 1 \pm \frac{\Phi_1}{2} - \frac{\Phi_2}{8} \pm \frac{\Phi_3}{16} - \ldots \right) \quad ,$$

$$\Phi_1 = \psi \left( \mu_o + \vartheta \right) \quad ,$$

$$\Phi_2 = \psi^2 \left( 3\mu_o^2 + 2\vartheta\mu_o - \vartheta^2 \right) \quad ,$$

$$\Phi_3 = \psi^3 \left( \vartheta^3 - \vartheta^2\mu_o + 3\vartheta\mu_o^2 + 5\mu_o^3 \right) \quad ,$$

$$\psi = \frac{e^2}{4\pi m r} \quad , \quad \vartheta = \frac{1}{\varepsilon_o \omega_o^2 r^2} \quad .$$

We now solve the problem of our coupled oscillators. Planck's original result involved counting the number of uncoupled harmonic oscillators partaking in the field's motions in a system averaged over some random phase $\phi$ at each frequency. If instead of non-interacting modes a system with a fixed number of oscillators has two coupled modes, the zero-point energy will vanish at first order. This point has also been made by Post. We assume that our oscillators can be diagonalized into two fundamental modes, of correlated and anticorrelated motions whose eigenfrequencies are $\omega_-$ and $\omega_+$, that a majority of the oscillators contribute to the $\omega_-$ mode, that an oscillator whose phase shifts from $\phi$ to $\phi+\pi$ increases the occupation number of one mode while decreasing that of the other, and that one such isolated transition does not substantially alter the fundamental frequency of the system.

$$[11] \quad \xi = \hat{q}_- \sin(\omega_- t + \phi_-) \quad , \quad \eta = \hat{q}_+ \sin(\omega_+ t + \phi_+) \quad .$$

The energy of a single coupling mode is

$$[12] \quad E_\xi = \frac{1}{2} k_- \hat{q}_-^2 \quad , \quad E_\eta = \frac{1}{2} k_+ \hat{q}_+^2 \quad ,$$

and that of a system bound between condition *I* and condition *II* is found by substituting [11] into [9], and then integrating over the collective phase-space after a change of coordinates $\Lambda$

$$[13]\quad \langle E\rangle = \frac{\iint\limits_{I\to II} \left(E_\xi + E_\eta\right) dp_\xi\, d\xi\, dp_\eta\, d\eta}{\iint\limits_{I\to II} dp_\xi\, d\xi\, dp_\eta\, d\eta} \quad ,$$

$$\Lambda(q_\pm, p_\pm) = (\hat{q}_\pm, \phi_\pm) \quad , \quad J(\Lambda) = m\hat{q}_\pm\omega_\pm \quad ,$$

$$\begin{aligned}\langle E\rangle &= \frac{k_+}{4}\frac{\hat{q}_+^{II4} - \hat{q}_+^{I4}}{\hat{q}_+^{II2} - \hat{q}_+^{I2}} + \frac{k_-}{4}\frac{\hat{q}_-^{II4} - \hat{q}_-^{I4}}{\hat{q}_-^{II2} - \hat{q}_-^{I2}} \\ &= \frac{k_+}{4}\left(\hat{q}_+^{I2} + \hat{q}_+^{II2}\right) + \frac{k_+}{4}\left(\hat{q}_-^{I2} + \hat{q}_-^{II2}\right) \\ &= \frac{1}{2}\left(E_I + E_{II}\right)\end{aligned}$$

Next, we write the energies corresponding to those bounding conditions for a system constrained by its fixed number of oscillators divided into a correlated mode that dominates the dynamics and its complement :

$$[14]\quad E_{I-} = \frac{k_-}{2}\hat{q}_-^{I2} = n_-\hbar\omega_- \quad , \quad E_{II-} = (n_- + 1)\hbar\omega_-$$

$$E_{I+} = \frac{k_+}{2}\hat{q}_+^{I2} = n_+\hbar\omega_+ \quad , \quad E_{II+} = (n_+ - 1)\hbar\omega_+$$

$$\langle E\rangle = \frac{1}{2}\left(E_I + E_{II}\right) = \hbar\omega_-\left(n_- + \frac{1}{2}\right) + \hbar\omega_+\left(n_+ - \frac{1}{2}\right) \quad .$$

$$N = n_+ + n_-$$

This describes the evolution of the constrained system under its dominant coherent mode whose occupation number increases by one quantum while the its complement is similarly depleted. The correlation is transitive. Its modes form coherent states in large ensembles weighted by their number $n_-$ whereas the anticorrelation self-destructs, as its modes decrease the collective strength of the overall interaction, including their own coupling energy to the ensemble. When the anticorrelated modes reach their maximum number, which is always less than half the total, the disordered system has no long-range interaction. So it is natural for the anticorrelated zero-modes to somewhat resemble those of a fermionic oscillator. This constraint is similar to that which applies in a two-fluid superconductor wherein $N_s = N - N_n$ , and the subscipts *s* and *n* denote the superconducting and normal phases. This explains why superconductors have a vanishing zero-point energy. Using [10], the mean energy can be approximated as

[15] $$\langle \boldsymbol{E} \rangle_\omega \cong \hbar\omega_{\boldsymbol{o}}\left( (\boldsymbol{n}_+ - \boldsymbol{n}_- - \mathbf{1})\left( \Phi_1 + \frac{\Phi_3}{\mathbf{16}} + \mathbf{...} \right) + (\boldsymbol{n}_+ + \boldsymbol{n}_-)\left( \mathbf{1} - \frac{\Phi_2}{\mathbf{8}} - \mathbf{...} \right) \right) \quad .$$

The upper bound of the spectrum of interest being within the near-field, the mean energy is

[16]

$$\langle \boldsymbol{E} \rangle = \frac{\int_{\omega = \boldsymbol{o}}^{\boldsymbol{c}/\boldsymbol{r}} \rho(\omega) \langle \boldsymbol{E} \rangle_\omega}{\int_{\omega = \boldsymbol{o}}^{\boldsymbol{c}/\boldsymbol{r}} \rho(\omega)} \quad ,$$

whose *1/r* dependence may only result from a spectrum that is strongly peaked in the infrared. A discrete distribution $\rho_\omega$ with such strongly peaked dominant frequencies is a natural consequence of coherence, which gives, for the zero modes and within a coherence domain,

[17]

$$\langle \boldsymbol{E} \rangle \cong \; -\hbar\langle \omega \rangle \left( \Phi_1 + \frac{\Phi_3}{\mathbf{16}} + \mathbf{...} \right) \quad ,$$

with no cosmological constant. If instead the system were disordered owing to too high a temperature ($n_+ \sim n_-$) or a limited number of oscillators, then the standard, unconstrained counting of the modes applies and

[18] $$\langle \boldsymbol{E} \rangle_\omega \cong \hbar\omega_{\boldsymbol{o}}\left( (\boldsymbol{n}_+ + \boldsymbol{n}_- + \mathbf{1})\left( \mathbf{1} - \frac{\Phi_2}{\mathbf{8}} + \mathbf{...} \right) + (\boldsymbol{n}_+ - \boldsymbol{n}_-)\left( \frac{\Phi_3}{\mathbf{16}} + \mathbf{...} \right) \right) \quad ,$$

with the following contribution for the zero-modes

[19]

$$\langle E\rangle_{\omega} \cong \hbar\langle\omega\rangle\left(1-\frac{\Phi_2}{8}+...\right) \quad ,$$

in which one recognizes the standard Van der Waals potential, whose longest-range component is electrokinetic and varies as $1/R^2$ , and whose dominant term in the short range is the purely coulombian $1/R^6$. In extended systems devoid of coherent resonances, the spectrum is continuous and [18] must be integrated over all the frequencies of the near field that appear in [16], which induces a positive and globally antigravitating cosmological constant, while the longest-range component varies as $1/R^3$, as pointed out by Cole et al. Returning to our subject, it is obvious that the usual electromagnetic polarization of matter cannot form coherent modes over macroscopic distances at room temperature or above, since the allowed coherence lengths are generally smaller than $10\mu$. On the other hand, a far infrared-dominated coherent spectrum that allows for [14]-[17] can arise naturally in Machian cosmologies, whose potentials only decrease as $1/r$, while the matter distribution is constant, as Sciama had pointed out. For instance, in Yilmaz's theory of gravitation, the far Universe, though at infinity, takes on the appearance of an asymptotically thin, dense and infinitely red-shifted shell sitting at a finite distance. In such a Euclidian Universe, a measuring rod $\Delta r$ transported some distance $r$ will stretch, owing to its new location being surrounded by a homogeneous sphere of matter density $\rho$. The distance that it measures appears shortened as

[20]

$$\Delta r' = e^{-\frac{M_{(r)}G}{rc^2}}\Delta r$$

$$= e^{-\frac{4\pi\rho\, G r^2}{3c^2}}\Delta r \quad ,$$

so that the measured radius of an unbounded Universe becomes

[21]

$$R' = \int_{o}^{\infty} e^{-\frac{4\pi\rho\, G r^2}{3c^2}}\, dr \quad ,$$

which is finite ($\sim 15 \times 10^9$ *LYrs*), while the red-shift $z_{(r)}$ of distant matter goes to infinity as

[22]

$$z = e^{\frac{4\pi\rho\, G r^2}{3c^2}} - 1 \quad ,$$

at a finite *R'*. The phase Φ of the photons in the coherent modes and hence their position being essentially uncertain because of a fixed *N* and $\omega_o$, an infinite red-shift at a finite effective distance causes the electronic and partonic Zitterbewegung frequencies of far away matter to acquire wavelengths larger than *R'*, so that its evanescent radiative field is felt here, in direct proportion to its energy. This is in agreement with the equivalence principle. How exactly is it felt ? Zitterbewegung involves oscillatory motion at the speed of light. For any observer at that speed, the photon bath is infinitely shifted to the blue. Zitterbewegung is oscillatory because at light speed, a charge ends up being back-scattered by such photons within a short interval of time. Their dense, far infrared virtual background is seemingly invisible to macroscopic observers at non ultra-relativistic speeds, save for its inertia-inducing effects. It is nonetheless felt by the charged oscillating microscopic constituents of matter, which propagate at light speed. This explains how momentum-energy far away induces inertia here. Being the main contributor to inertia, the effective shell of the faraway Universe plays a similar role in cohering gravity by the spectral density ρ(ω) of its coherent modes in [16]. The anticorrelated modes are obviously the negative energy electronic states of Dirac's theory. Equating the gravitational to the oscillator energy in *1/r* between two electrons gives $\nu_o = 10^{-22} Hz$. Relative to the electronic Zitterbewegung frequency, this represents a reddening by $6x10^{-43}$, and a distance slightly larger than that beyond which the Zitterbewegung of distant electrons can be felt, which is at about eleven times the *apparent* radius of the universe with a reddening by $\sim 10^{-40}$. But $\nu_o$ is an eigenfrequency of our coupled oscillators and it is implausibly low as well as seemingly external to their dynamics. The solution is to consider a closed Universe, that is, we take $\Omega = 1+\varepsilon$ and either identify the antipodes after a parity reversal, or consider a tunneling over the extent of the hypersphere at a given time. In either case, it is the red-shifted radiative image of the local matter that appears spread here in the far infrared while the tunneling is assumed to be instantaneous from here to here, through a distance. In other words, the energy borrowed from the vacuum by virtual photons of energy $\sim mc^2$ is entirely returned by their cosmological redshift, the process being coherent. The scalar Riemannian curvature *R* thus appears in the one-loop diagrams of QFT while the Newtonian force emerges only at the two-loop level. Now the range of the near field of our coupled oscillators reaches to the Universe, within which coherent modes can arise in which the local oscillators fully partake if they are assumed to lie entirely within their coherence domain. The equivalence principle also emerges naturally while the critical temperature at which the coherence and therefore the usual gravitational interactions vanish for a particle of mass $m$ is at $kT \sim mc^2$, in accordance with the results of Donoghue et al. The red-shifted radiative image of local matter, through the magnifying glass of distance, partakes in global coherent modes, in the emergence of the local vacuum and waves the dynamic tapestry of space. In so doing, it binds to matter beyond the range of its generally assumed near field, inducing the $1/r$ gravitational potential. Since it is the kinetic motions of matter that induce gravity, the fall into a gravitational potential well reduces the fundamental frequencies of free matter, slows clocks down and decreases the mass-energy asymptotically.

The inductive coupling takes place in the near field $\nu_{\pm} << c/r_{12}$ between coherently coupled individual dipoles, through their red-shifted local antipodal image. This allows the exchanged photons to be virtual and the coherent modes to genuinely belong to the coupled oscillators while ensuring that the range of gravity spans the Universe. In this sense, the Zitterbewegung of all matter near and far can be felt here, in the far infrared ($\lambda_{red\ shifted} > 10^{10}\ LYrs$), by the Zitterbewegung of a test charged particle or dipole, in direct proportion to the rate of both, that is, to their energy, owing to their common coherent modes with the universe at large, through red-shifted tunneling photons. This is in agreement with the equivalence principle. Being thus cohered from the far infrared, the standard $1/r$ gravitational potential never vanishes so long as the coherent modes exist. The gradual increase of the excited anticorrelated states over their correlated counterparts with increasing temperature creates an antigravitational effect whose critical temperature is $T \sim mc^2/k$, the thermal random motions tending to destroy the correlated order. This result emerges in the finite temperature QED computations of Donoghue et al. In cosmology, the interacting modes no longer extend just about the oscillator in some finite near-field volume that shrinks as the cube of the frequency in its immediate surroundings, as would be expected, but reach in the far infrared to the dynamics of the most distant matter. As a result, the local Zitterbewegung becomes coherent. Induced gravity from electromagnetism is well suited to unified electromagnetic models of matter and to Dyonic schemes, such as those A.O. Barut put forth. The local inertial reaction relative to the past light-cone can be attributed to an active interaction with the radiative image of its field, material and electromagnetic, which can be likened to a classically coherent aether, as opposed to its incoherent ZPF counterpart that induces the Casimir and the Van der Waals forces. Athough the tunneling is instantaneous, it is not clear whether the effective local interactions are. The main difference with the standard ZPF approach is that the past light cone's inertia-inducing fluctuations are dominated by the very low tunneling frequencies that originate in the most distant matter of the Universe and that form coherent modes. Another difference is that it is the dynamics of the various states within matter itself, including that of its zero-modes, that induces the *1/r* attraction.

To conclude, the standard procedure for deriving the Van der Waals forces, along with an electrokinetic coupling, yields a *1/r* attractive potential between coherently coupled oscillating dipoles. In this framework, gravitation is an emergent phenomenon that is induced by the coherent Zitterbewegung of charges and its coupling to that of distant matter through the dilated and red-shifted components of their radiative field. We made one step towards the strong and weak equivalence principles, as all mass-energy, inertia and gravity are fundamentally in proportion to the Zitterbewegung of both local and distant charges and dipoles. At close range, the omitted Lorentz force is felt, which requires going to two or three spatial dimensions. This was done in a companion paper. The magnetic field is genuinely confined in the near field in its accepted sense and cannot affect the tunneling photons that are relevant on cosmological scales. In the long range, the electrokinetic interaction dominates, so that this simple model can be considered. If we were to assign each $1/r^N$ force to gravitons propagating in the bulk of an $N+1$-dimensional space including our 3-brane, then the standard Van der Waals force could be seen as arising from a space-time of nine dimensions !